\documentclass[superscriptaddress,aps,prd,nofootinbib,preprint]{revtex4-1}
\pdfoutput=1

\usepackage{graphicx}
\usepackage[colorlinks,allcolors=blue]{hyperref} 
\usepackage{amsmath,amssymb,bm}
\usepackage[T1]{fontenc}
\usepackage{mathtools}
\usepackage{physics}

\begin{document}

\preprint{TU-1273}

\title{Sign-Flipping Axion Potentials via Kapitza-Type Modulation by Heavy Axions}

\author{Kai Murai}
\email{kai.murai.e2@tohoku.ac.jp}
\affiliation{Department of Physics, Tohoku University, Sendai, Miyagi 980-8578, Japan}
\author{Fuminobu Takahashi}
\email{fumi@tohoku.ac.jp}
\affiliation{Department of Physics, Tohoku University, Sendai, Miyagi 980-8578, Japan}
\affiliation{Kavli IPMU (WPI), UTIAS, University of Tokyo, Kashiwa 277-8583, Japan}
\author{Kosei Tomiyama}
\email{tomiyama.kousei.p2@dc.tohoku.ac.jp}
\affiliation{Department of Physics, Tohoku University, Sendai, Miyagi 980-8578, Japan}

\begin{abstract}
We show that the potential of a light axion can flip sign, or even nearly vanish, as a result of coherent oscillations of a heavier axion with which it mixes.
This phenomenon is analogous to the Kapitza pendulum, where a high-frequency external force stabilizes an otherwise unstable configuration,
but here it arises naturally from the inherent mass hierarchy and mixing among axions in the axiverse, without the need for any externally imposed modulation. We further show that a late-time sign flip of the potential can significantly enhance the abundance of the light axion, which has important cosmological and observational consequences.
\end{abstract}

\maketitle

\section{Introduction}
\label{sec: intro}
Axions are strongly motivated by extensions of the Standard Model~\cite{Peccei:1977hh,Peccei:1977ur,Weinberg:1977ma,Wilczek:1977pj} and are among the leading candidates for dark matter (see Refs.~\cite{Kim:2008hd,Arias:2012az,Marsh:2015xka,DiLuzio:2020wdo,OHare:2024nmr} for reviews). In string theory, a large number of axion fields naturally appear through compactification, and they generally exhibit both kinetic and mass mixing. Their masses are determined by non-perturbative effects associated with different topological cycles in the compactification manifold. As a result, the axion mass spectrum often spans many orders of magnitude. Such a multitude of axions is referred to as the axiverse~\cite{Arvanitaki:2009fg} or the axion landscape~\cite{Higaki:2014pja}.

Mixing among multiple axions gives rise to a wide variety of cosmological phenomena. Representative examples include a QCD axion whose coupling to photons or dark photons is clockwork-enhanced~\cite{Higaki:2016yqk,Farina:2016tgd}, level-crossing events between the QCD axion and an axion-like particle~\cite{Hill:1988bu,Kitajima:2014xla,Daido:2015bva,Daido:2015cba,Ho:2018qur,Murai:2023xjn,Cyncynates:2023esj,Li:2023xkn,Li:2023uvt,Li:2024psa,Li:2024okl,Murai:2024nsp,Dunsky:2025sgz}, axion-driven inflation~\cite{Choi:2014rja,Higaki:2014pja,Nakayama:2014hga,Daido:2017wwb,Takahashi:2019pqf,Kobayashi:2019eyg,Kobayashi:2020ryx,Narita:2023naj},  resonance through non-linear interactions~\cite{Cyncynates:2021xzw}, misalignment mechanism~\cite{Murai:2023xjn}, decay of axions into Standard Model particles through mixing~\cite{Higaki:2014qua}, and the formation of complex topological defects that consist of axion strings and domain walls~\cite{Higaki:2016jjh,Long:2018nsl,Eto:2023aqr,Lee:2024xjb,Lee:2024toz,Lee:2025zpn,Kondo:2025hdc}. These phenomena illustrate the rich and diverse dynamics inherent in multi-axion systems.

One important aspect of axion dynamics is the timing at which the field begins to oscillate. This typically occurs when the curvature of the axion potential, or equivalently its mass, becomes comparable to the Hubble parameter. When the axion field starts near the top of its potential, however, the onset of oscillations can be delayed due to anharmonic effects~\cite{Kobayashi:2013nva}. This delay can enhance isocurvature perturbations and may also lead to spatial instabilities that result in the formation of localized structures such as oscillons~\cite{Bogolyubsky:1976yu,Gleiser:1993pt,Kasuya:2002zs}. In recent years, several new scenarios have been proposed in which the oscillation of the axion is significantly postponed. These include the trapped misalignment mechanism~\cite{Higaki:2016yqk,Nakagawa:2020zjr,DiLuzio:2021gos,Jeong:2022kdr,DiLuzio:2024fyt}, where the axion remains temporarily in a false vacuum, and the bubble misalignment scenario~\cite{Nakagawa:2022wwm,Lee:2024oaz}, in which the axion mass is suddenly generated by a first-order phase transition.

In this paper, we focus on a system of two axions with a large mass hierarchy. A heavy axion mixes with a light axion and undergoes coherent oscillations. Unlike in level-crossing scenarios, the masses remain well separated. At first glance, one might attempt to integrate out the heavy axion by fixing it at the minimum of its potential, thereby obtaining an effective potential for the light axion. We show, however, that this approach fails when the oscillation amplitude of the heavy axion is large. Such a situation can be easily realized in the trapped, bubble misalignment, or clockwork scenarios. After taking a proper time average over the rapid oscillations, we find that the low-energy effective potential of the light axion can flip sign,%
\footnote{A sign flip of the axion potential is also discussed in the context of extremely dense environments such as neutron stars~\cite{Hook:2017psm,Balkin:2020dsr,Balkin:2023xtr,Kumamoto:2024wjd} or extensions of the QCD axion model~\cite{Daido:2017wwb,Co:2018mho,Takahashi:2019qmh,Takahashi:2019pqf,Nakagawa:2020eeg,Huang:2020etx,Gonzalez:2022mcx,Narita:2023naj,Co:2024bme,Khoury:2025txd}, though the underlying mechanisms in those cases are different from the dynamical modulation considered in this work.}
and even vanish nearly completely
at certain oscillation amplitudes of the heavy axion.
We also analyze the resulting spatial instabilities in this regime, as well as the impact of a late-time sign flip on the axion abundance.

The striking behavior of this rapidly oscillating system closely parallels the case of Kapitza's inverted pendulum~\cite{Kapitza1951,Stephenson1908}, in which a high-frequency drive stabilizes an otherwise unstable configuration. Kapitza-type modulation, or more broadly Floquet modulation, is widely used in condensed matter physics and quantum devices~\cite{Bukov2015}. In our setup, however, the modulation is not imposed externally, but arises dynamically from the coherent oscillations of a heavy axion through mixing, with both ingredients naturally appearing in the axiverse. To the best of our knowledge, this work presents the first explicit application of a Kapitza-type mechanism in axion physics and explores its cosmological implications.

The rest of this paper is organized as follows.
In Sec.~\ref{sec: background}, we introduce the setup of the two-axion system and explain the Kapitza-type modulation of the axion potential.
We then demonstrate the validity of the time averaging and speculate on specific cases where the effective description breaks down.
In Sec.~\ref{sec: particle production}, we analyze perturbations of the light axion and their growth.
We discuss cosmological applications of the Kapitza-type modulation and evaluate the enhancement of the axion abundance in Sec.~\ref{sec: app}.
Section~\ref{sec: summary} is devoted to the summary and discussion of our results.

\section{Axion-Induced Kapitza-Type Modulation}
\label{sec: background}

\subsection{Physical Setup}
\label{subsec: model}

Let us consider a two-axion system with mixing and hierarchical masses, as motivated by the axiverse. 
We assume that the heavy axion $\phi$ is spatially homogeneous. While we initially consider the light axion $\chi$ with spatial dependence when deriving the equations of motion, in this section, we focus on the evolution of its homogeneous component. The growth of fluctuations will be discussed in the next section. 

The heavy axion $\phi$ has a periodic potential given by 
\begin{align}
    V_\phi(\phi)
    =
    m_\phi^2 f_\phi^2 
    \left[
        1 - \cos\left( \frac{\phi}{f_\phi} \right) 
    \right],
    \label{potential0}
\end{align}
where $m_\phi$ and $f_\phi$ are the mass and decay constant of $\phi$.
To introduce a mixing between the two axions, we consider the following potential:
\begin{align}
    V_\mathrm{mix}(\phi,\chi)
    =
    m_\chi^2f_\chi^2
    \left[ 
        1 - \cos\left( N\frac{\phi}{f_\phi}+\frac{\chi}{f_\chi} \right)
    \right],
\label{mixing potential}
\end{align} 
where 
$m_\chi$ approximately corresponds to the mass of $\chi$ after integrating out the heavy axion $\phi$,
and $f_\chi$ is the decay constant of $\chi$.
Here, we define decay constants such that $V_\phi$ is periodic under $\phi \to \phi + 2\pi f_\phi$ and the mixing potential $V_\mathrm{mix}$ is periodic under $\chi \to \chi + 2\pi f_\chi$.
Under this convention, the parameter $N$ is generically a rational number rather than an integer. We assume hierarchy in the axion masses, $m_\phi \gg m_\chi$.

In this paper, we focus on the case where the backreaction of the light axion $\chi$ on the heavy axion can be neglected, so that the heavy axion can be treated as an external field.
The equation of motion for the heavy axion $\phi$ is
\begin{align}
\ddot{\phi} + m_\phi^2 f_\phi \sin\left(\frac{\phi}{f_\phi}\right)
+ m_\chi^2 \frac{f_\chi^2}{f_\phi} N \sin\left(N\frac{\phi}{f_\phi} + \frac{\chi}{f_\chi}\right)
&= 0,
\label{eq: full heavy}
\end{align}
where the second and third terms correspond to $\partial V_\phi/\partial \phi$ and $\partial V_{\rm mix}/\partial \phi$, respectively.
For the backreaction to be negligible, the contribution from $V_{\rm mix}$ must remain much smaller than that from $V_\phi$.
As we will see below, for the Kapitza effect to be relevant, the argument of the sine in the $\partial V_{\rm mix}/\partial \phi$ term should be larger than ${\cal O}(1)$, while the oscillation amplitude of $\phi$ is generically smaller than $f_\phi$.
Thus, the condition can be written as
\begin{align}
m_\phi^2 f_\phi \Phi \gg N m_\chi^2 f_\chi^2,
\label{eq: bacreaction}
\end{align}
where $\Phi$ denotes the oscillation amplitude of $\phi$. We will return to this condition later when evaluating the abundance of $\phi$ in the expanding universe.

\subsection{Effective Potential via Time Averaging}
\label{subsec: effective}

In this section, we neglect cosmic expansion and work in the Minkowski spacetime here, and we will return to the case with cosmic expansion later.
The axions satisfy the equations of motion,
\begin{align}
    \ddot{\phi} + m_\phi^2 f_\phi \sin \left(\frac{\phi}{f_\phi}\right)
    &\simeq 
    0,
    \label{eq: heavy}
    \\
    \ddot{\chi} - \nabla^2\chi + m_\chi^2 f_\chi \sin\left( N\frac{\phi}{f_\phi} + \frac{\chi}{f_\chi} \right)
    &=
    0,
    \label{eq: light}
\end{align}
where we have neglected the gradient of $V_\mathrm{mix}$ in the equation of motion for $\phi$ because of the hierarchy between $V_\phi$ and $V_\mathrm{mix}$.
 
When the amplitude of $\phi$ is smaller than $f_\phi$, the potential can be approximated as $V_\phi \simeq\frac{1}{2}m_\phi^2\phi^2$, and the time evolution of $\phi$ is then given by
\begin{align}
    \phi(t)
    =
    \Phi\cos\left(m_\phi t+\alpha\right),
    \label{eq: phimotion}
\end{align}
where $\Phi$ is the oscillation amplitude, and $\alpha$ is a constant phase of oscillations.
When the amplitude $\Phi$ is comparable to or larger than $f_\phi$, the actual angular frequency becomes slightly smaller than $m_\phi$ due to the flatter shape of the potential than the quadratic one.
In the following, we assume the small amplitude of $\phi$ and use the solution~\eqref{eq: phimotion} for analytic estimates.
However, the essential point is that $\phi$ oscillates rapidly, and the following discussion remains qualitatively valid as long as $\phi$ oscillates more rapidly than $\chi$, although the analytic expressions and quantitative details may be different.
In our numerical analysis, we do not restrict ourselves to the small-amplitude regime, and the same qualitative behavior is observed.

In the following, we treat $\phi$ in Eq.~\eqref{eq: phimotion} as an external field and investigate the evolution of $\chi$ in the background of $\phi$.
Since Eq.~\eqref{eq: light} is non-linear, its analytical treatment is in general challenging.
So, we decompose $\chi$ into a spatially homogeneous component $\bar{\chi}(t)$ and perturbations $\delta\chi(t,\bm{x})$ as
\begin{align}
    \chi(t,\bm{x}) = \bar{\chi}(t)+\delta\chi(t,\bm{x}).
\end{align}
For the moment we focus on the dynamics of the background field $\bar{\chi}$.
The equation of motion for $\bar{\chi}$ is given by 
\begin{align}
    \ddot{\bar{\chi}} + 
    m_\chi^2
    f_\chi \sin\left( A\cos\left(m_\phi t+\alpha\right)+ \frac{\bar{\chi}}{f_\chi} \right)
    =
    0,
    \label{eq: full}
\end{align}
where $A \equiv N \Phi/f_\phi$.
This equation of motion arises from the time-dependent  potential
\begin{align}
    V_\mathrm{mix}(\bar{\chi},t)
    =
    m_\chi^2
    f_\chi^2\left( 1 - \cos ( A \cos \left(m_\phi t+\alpha\right) + \frac{\bar{\chi}}{f_\chi} )\right)
    \ ,
\end{align}
where we have substituted Eq.~\eqref{eq: phimotion}.

\begin{figure}[t]
    \centering
    \begin{minipage}[t]{0.48\linewidth}
        \centering
        \includegraphics[width=\linewidth]{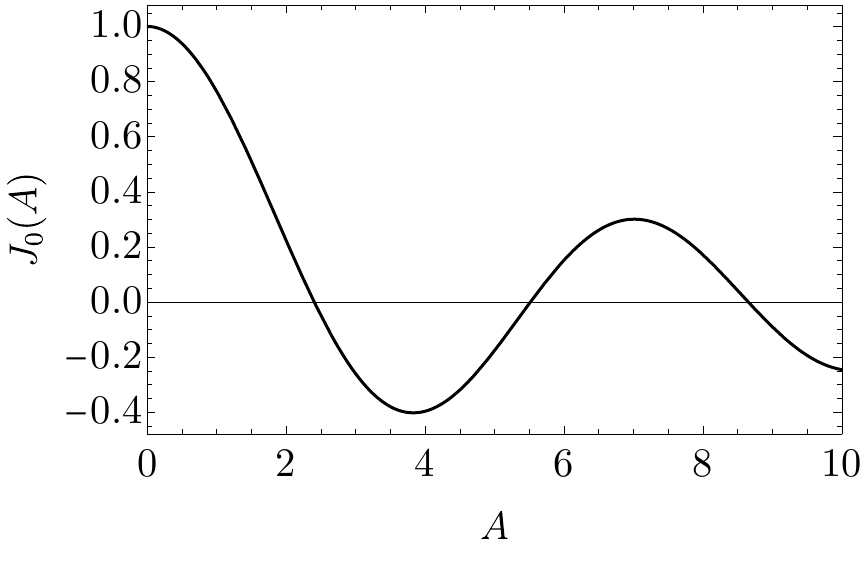}
        (a) Bessel function $J_0(A)$
    \end{minipage}
    \hfill
    \begin{minipage}[t]{0.5\linewidth}
        \centering
        \includegraphics[width=\linewidth]{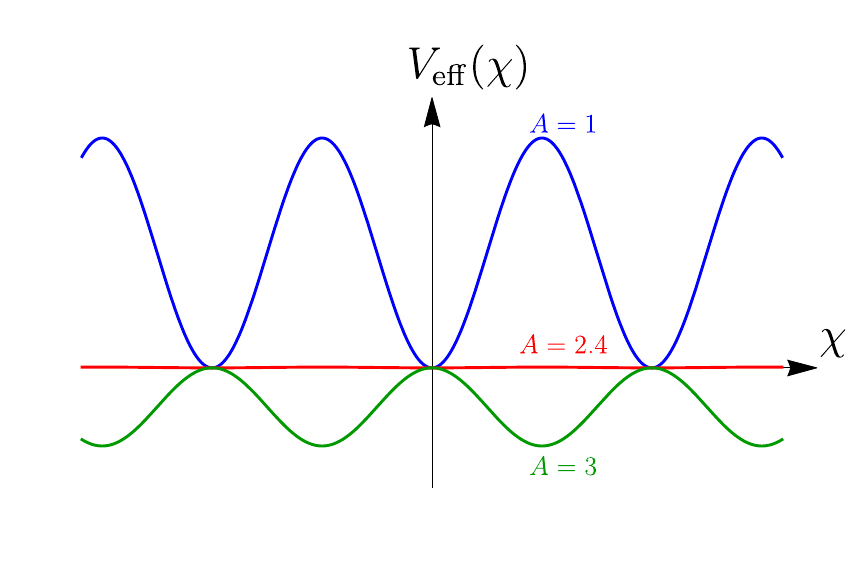}
        (b) Effective potential
    \end{minipage}
    \caption{%
    \textit{Left}: Bessel function $J_0(A)$, which has zeros at $A \simeq 2.4$, $5.5$, $8.7$, $\ldots$.
    \textit{Right}: Effective potential~\eqref{eq: effective potential}, whose height and stable points depend on the value of $A$.}
    \label{fig: J0}
\end{figure}

While $\phi$ oscillates with an angular frequency of $m_\phi$, the characteristic timescale of $\bar{\chi}$ is set by $m_\chi$.
Since we consider the regime $m_\phi \gg m_\chi$, $\bar{\chi}$ evolves much more slowly than $\phi$.
Thus, we expect that the evolution of $\bar{\chi}$ can be described by an effective equation of motion, which is obtained by averaging over the rapid oscillations of $\phi$.
To see this, we expand the potential using the addition formula,
\begin{align}
    &\cos\left[A\cos\left(m_\phi t+\alpha\right)+\frac{\bar{\chi}}{f_\chi}\right] \notag \\
    &=
    \cos\left[A\cos(m_\phi t+\alpha)\right]\cos\left(\frac{\bar{\chi}}{f_\chi}\right)
    -
    \sin\left[A\cos(m_\phi t+\alpha)\right]\sin\left(\frac{\bar{\chi}}{f_\chi}\right)
    ,
\end{align}
and then average over one period of $\phi$ oscillations as 
\begin{align}
   \frac{m_\phi}{2\pi} \int_0^{\frac{2\pi}{m_\phi}} \mathrm{d} t \,
    \cos[ A \cos (m_\phi t + \alpha) ]
    &=
    J_0(A)
    \ , \\
     \frac{m_\phi}{2\pi} \int_0^{\frac{2\pi}{m_\phi}} \mathrm{d} t \,
    \sin[ A \cos (m_\phi t + \alpha) ]
    &=
    0  
    \ ,
\end{align}
where $J_0(A)$ is the Bessel function of the first kind, which is shown in the left panel of Fig.~\ref{fig: J0}.
As a result, the effective potential is given by 
\begin{align}
    V_\mathrm{eff}(\bar{\chi})
    =
    m_\chi^2
    f_\chi^2J_0(A)\left[1-\cos \left(\frac{\bar{\chi}}{f_\chi}\right) \right],
    \label{eq: effective potential}
\end{align}
where we shifted the constant term so that $V_\mathrm{eff}(0) = 0$.
Then, the equation of motion \eqref{eq: full} is simplified to
\begin{align}
    \ddot{\bar{\chi}}+
    m_\chi^2
    f_\chi J_0(A)\sin\left(\frac{\bar{\chi}}{f_\chi}\right)
    =
    0
    \label{eq: timeaverage}.
\end{align}
Note that this time averaging becomes invalid when the characteristic time scales of $\phi$ and $\bar{\chi}$ are not sufficiently separated. 

We emphasize that the effective potential depends on the oscillation amplitude of the heavy axion through the Bessel function $J_0(A)$. As shown in Fig.~\ref{fig: J0}, $J_0(A)$ can take 
positive or negative values, depending on $A$. In particular, 
a sign flip interchanges its maxima and minima: when $J_0(A) > 0$, $\bar{\chi} = 0$ is the stable minimum, while for $J_0(A) < 0$, $\bar{\chi} = 0$ becomes an unstable maximum and $\bar{\chi} = \pi f_\chi$ turns into the stable minimum. This behavior is in contrast to $V_\mathrm{mix}$ with $\phi$ fixed at its oscillation center or the low-energy minimum ($\phi = 0$).
This provides a realization of Kapitza-type modulation in axion dynamics, where rapid oscillations of the phase modify not only the height but also the sign of the effective potential.


\subsection{Numerical Demonstration of the Kapitza Effect}

Here we present numerical results illustrating the emergence of the Kapitza effect through time averaging.
For this purpose, we introduce the dimensionless variables $\theta_\phi \equiv \phi/f_\phi$ and $\theta_\chi \equiv \bar{\chi}/f_\chi$,  and use $m_\chi t$ as the dimensionless time.
We also parameterize the mass hierarchy by $R_m \equiv m_\chi/m_\phi < 1$.

\begin{figure}[t!]
    \begin{center}
    \includegraphics[width=0.8\linewidth]{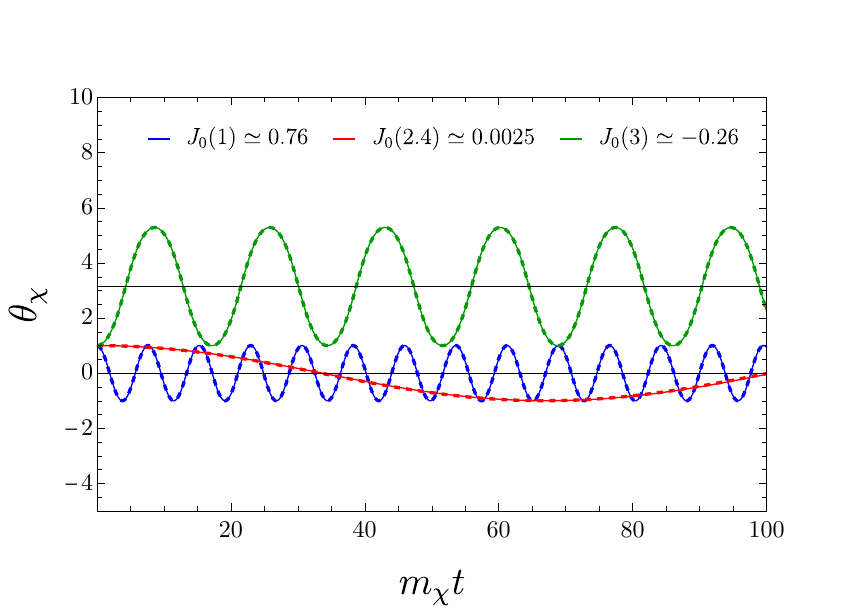}
    \end{center}
    \caption{Numerical solutions of the original (solid) and time-averaged (dashed) equations of motion for $\theta_\chi = \bar{\chi}/f_\chi$, with parameters $R_m = 0.01$, $\theta_\chi(0) = 1$, and $\dot{\theta}_\chi(0) = 0$. 
    The blue, red, and green lines correspond to $A=1$, $2.4$, and $3$, respectively.
    The horizontal black lines represent $\theta_\chi = 0$ and $\pi$.
    The time-averaged dynamics closely reproduces the original dynamics.}
    \label{fig: coin}
\end{figure}
In Fig.~\ref{fig: coin}, we show the solutions of the original equation of motion~\eqref{eq: full} (solid lines) and the time-averaged equation of motion~\eqref{eq: timeaverage} (dashed lines) for different values of $A$.
Here, we take $\alpha = 0$ and $R_m = 0.01$, and set the initial condition as $\theta_\chi(0) = 1$ and $\dot{\theta}_\chi(0) = 0$.
We find that the solution of the original equation~\eqref{eq: full} is well reproduced by the time-averaged dynamics.
We also confirm that the solution of Eq.~\eqref{eq: full} shows almost no dependence on $\alpha$, which is consistent with the fact that Eq.~\eqref{eq: timeaverage} becomes independent of $\alpha$ after time averaging.
We can see that the oscillation period for $A = 2.4$ (red) is longer than that for $A = 1$ (blue) due to the difference in the height of the effective potential, or equivalently, $J_0(A)$.
Furthermore, for $A = 3$, the light axion oscillates around $\theta_\chi = \pi$, which is an unstable point of $V_{\rm mix}$ in the absence of the heavy axion $\phi$.
This is because the overall sign of the effective potential is flipped with $J_0(3)<0$. It is remarkable that these three qualitatively different behaviors arise for the same $m_\chi$, driven solely by the difference in the oscillation amplitude of the heavy axion with which it mixes.

\begin{figure}[t]
    \centering
    \includegraphics[width=0.8\linewidth]{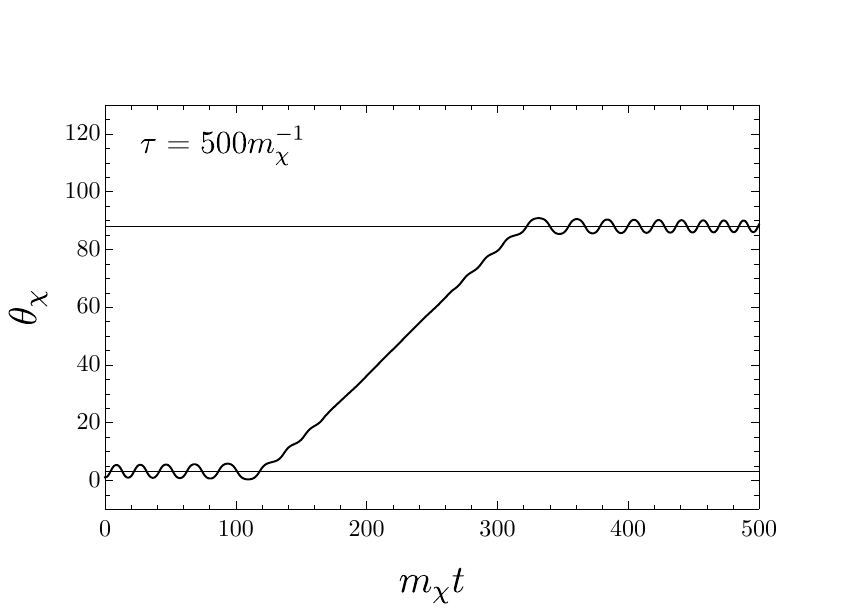}
    \caption{Time evolution of $\theta_\chi$ with a decreasing amplitude of the heavy axion.
    Here, we set 
     $A_\mathrm{in} = 3$, $\tau = 500 m_\chi^{-1}$,
    $R_m=0.01$, $\theta_\chi\left(0\right) = 1$, and $\dot{\theta}_\chi\left(0\right)=0$.
    During the first oscillations, the Bessel function is negative, $J_0(A)<0$, and $\bar{\chi}$ oscillates around $\pi f_\chi$. 
    In the second oscillations after $m_\chi t\simeq 300$, $J_0(A)$ becomes positive,
    and $\bar{\chi}$ oscillates around $14 \times 2\pi f_\chi$.
    The horizontal lines represent $\theta_\chi = \pi$ and $28\pi$.
}
    \label{fig: flip}
\end{figure}

Next, we consider the case where $A$ (or $\Phi$) varies over time.
We assume that the amplitude of the heavy axion decreases due to axion decay or other damping processes.
As a result, the sign of the Bessel function $J_0(A)$ can flip, leading to an exchange of the stable and unstable points.
A detailed analysis of the case with cosmic expansion will be presented in Sec.~\ref{sec: app}.
Here, we focus on the time variation of $A$ and parameterize the decay of the amplitude of $\phi$ as
\begin{align}
    \Phi(t)
    =
    \Phi_\mathrm{in}\exp\left(-\frac{t}{2\tau}\right).
\end{align}
Then, $\bar{\chi}$ obeys the equation of motion,
\begin{align}
    \ddot{\bar{\chi}} + R_m^2 m_\phi^2 f_\chi \sin\left(
        A_\mathrm{in} \exp\left( -\frac{t}{2\tau} \right)\cos(m_\phi t)
        +\frac{\bar{\chi}}{f_\chi}
    \right)
    =
    0,
\end{align}
where $A_\mathrm{in} \equiv N \Phi_\mathrm{in}/f_\phi$ and we set $\alpha=0$.
The result of numerical calculation is shown in Fig.~\ref{fig: flip}. 
Initially, the Bessel function has a negative value $J_0(A)<0$, and $\bar{\chi}$ oscillates around $\pi f_\chi$, which corresponds to one of the stable points of the effective potential in the early stage of evolution.
As time progresses, $J_0(A)$ gradually approaches zero, and the effective potential becomes nearly flat.
During this phase, $\bar{\chi}$ undergoes free motion.
Subsequently, $J_0(A)$ increases and becomes positive, causing the potential to reappear and grow, which drives $\bar{\chi}$ to resume oscillations around $\theta_\chi = 14 \times 2\pi$.
It is difficult to understand such behavior without the effective potential.
The final oscillation center depends on parameters such as $\tau$ and the initial conditions.
After a sufficiently long time, the oscillation amplitude of the heavy axion effectively vanishes  $(A \to 0)$, so the center of the $\theta_\chi$ oscillation settles at $2\pi n$ $\left(n \in\mathbb{Z}\right)$. 

\subsection{Breakdown of the Effective Description}

We have confirmed the Kapitza-type effect in axion mixing by performing a time averaging over the rapid oscillations of the heavy axion $\phi$.
However, the effective description is valid only when the typical dynamical timescale of $\chi$ is much longer than the oscillation period of $\phi$, i.e., when $\chi$ evolves sufficiently slowly compared to the rapid modulation by $\phi$.
Here, we present two examples where this approximation breaks down.

\subsubsection{Insufficient Mass Hierarchy}

\begin{figure}[t]
    \centering
    \includegraphics[width=0.8\linewidth]{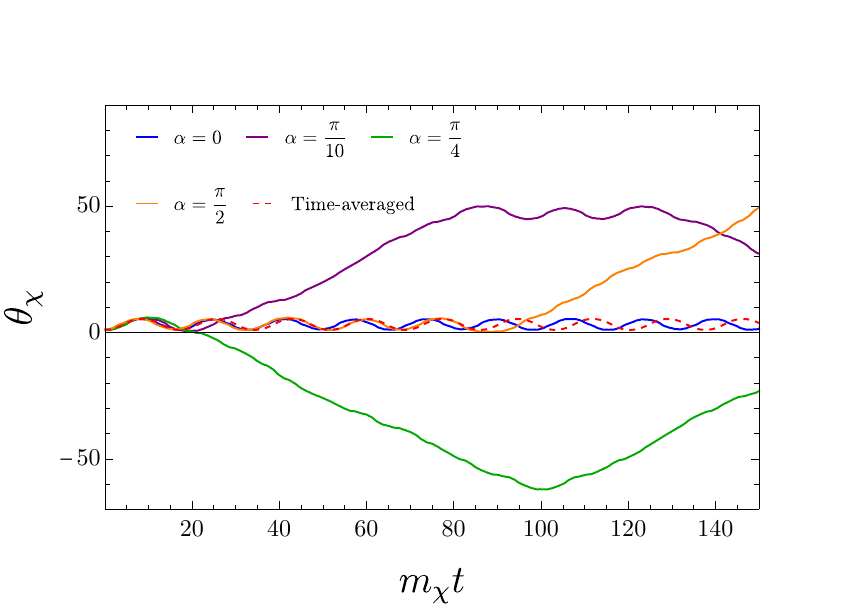}
    \caption{Time evolution of $\theta_\chi$ with $A=3$, $R_m = 0.4$, $\theta_\chi (0) = 1$, and $\dot{\theta}_\chi (0) = 0$.
    Due to the insufficient mass hierarchy,} the effective potential approximation fails, and the results depend strongly on the phase $\alpha$.     
    \label{fig: discrepancy}
\end{figure}

First, we consider the case in which the mass hierarchy is insufficient.
Fig.~\ref{fig: discrepancy} shows the time evolution of $\theta_\chi$ for $R_m = 0.4$.
In this case, the mass hierarchy between the two axions is not large enough, so the oscillation of $\phi$ significantly affects the evolution of $\bar{\chi}$,  and the use of the time-averaged equation is no longer valid.
As a result, as seen in the figure, the solutions to the original equation of motion (solid) and the time-averaged one (dashed) exhibit a significant discrepancy.
Moreover, the evolution of $\bar{\chi}$ becomes sensitive to the phase $\alpha$ of the $\phi$ oscillation, and changing $\alpha$ leads to substantial differences in the subsequent dynamics.
Since the value of the Bessel function depends on $A$, the range of $R_m$ for which the time-averaged description remains valid also changes accordingly.
In practice, the approximation typically breaks down around $R_m = \mathcal{O}(0.1)$ unless the field value of $\bar{\chi}$ is close to the hilltop.

\subsubsection{Hilltop Initial Conditions}

Next, we consider a situation in which $\bar{\chi}$ is initially located near the hilltop of the effective potential, e.g., $|\bar{\chi}(0)| \ll f_\chi$ when $J_0(A) < 0$.
In this case, $\bar{\chi}$ initially evolves slowly with a tiny amplitude, similar to slow-roll behavior.
When $\theta_\chi$ is sufficiently small, $\theta_\chi \ll 1$, 
Eq.~\eqref{eq: full} can be approximated as
\begin{align}
    \ddot{\bar{\chi}}+R_m^2
    m_\phi^2 f_\chi\left[
    \sin\left(A\cos\left(m_\phi t+\alpha\right)\right)
    + 
     \cos\left(A\cos\left(m_\phi t+\alpha\right)\right) \theta_\chi
    \right]
   \; \simeq\;
    0.
\end{align}
Although the second term on the left-hand side vanishes under time averaging, a small residual contribution generally remains due to the finite mass hierarchy.
On the other hand, the third term on the left-hand side gives rise to the Bessel function after time averaging, and it determines the effective potential discussed earlier.
However, for hilltop initial conditions with $|\theta_\chi| \ll 1$, this contribution becomes very small, and can become comparable to the residual from the second term.
In such cases, the motion of $\bar{\chi}$ may exhibit small fluctuations that are not captured by the effective potential, and these fluctuations can potentially have a significant impact on its dynamics.

\begin{figure}[t]
    \centering
    \includegraphics[width=1\linewidth]{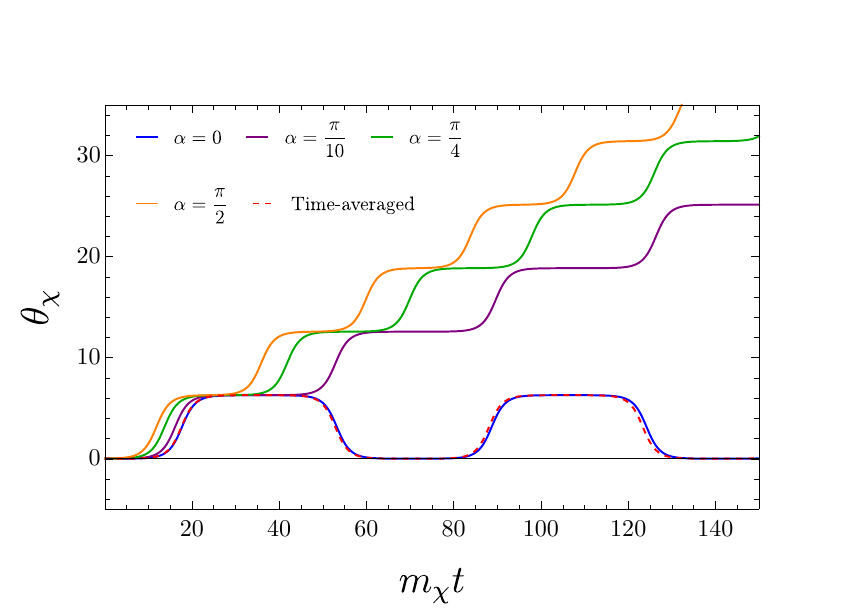}
    \caption{Time evolution of $\theta_\chi$ with $A=3$, $R_m=0.01$, $\theta_\chi (0) = 0.001$, and $\dot{\theta}_\chi (0) = 0$.
     The setup is the same as the green line in Fig.~\ref{fig: coin} except for $\theta_\chi (0)$ and $\alpha$.}
    \label{fig: hilltop}
\end{figure}
Fig.~\ref{fig: hilltop} shows the evolution of $\theta_\chi$ under the same conditions as the green line in Fig.~\ref{fig: coin}, except that the initial condition is set near the hilltop, with $\bar{\chi}(0)/f_\chi = 0.001$.
When $\bar{\chi}$ starts near the hilltop, the contribution of the effective potential to the equation of motion for $\bar{\chi}$ becomes extremely small.
As a result, the residual effects from the time-averaged oscillations of $\phi$ can become significant and influence the dynamics of $\bar{\chi}$.
Consequently, the solution becomes sensitive to the phase $\alpha$ of the $\phi$ oscillation.
In particular, we observe that the field remains near the hilltop for an extended period, and due to the residual fluctuations,  it eventually passes through a minimum and climbs up to the next maximum of the effective potential, where it stays for a long  again.
If the effective potential provided a perfect description, the field would simply oscillate between two adjacent maxima and never cross over to the next one (see the dashed line).
However, this numerical result shows that the residual fluctuations from the $\phi$ oscillation can drive the field across the potential barrier, allowing it to reach the next maximum.
The range of initial conditions that allow for such behavior clearly depends on the mass hierarchy between the two axions.
For a larger hierarchy, the residual becomes smaller after time averaging, and the initial condition must be set increasingly close to the hilltop in order for this phenomenon to occur.
Furthermore, Fig. \ref{fig: hilltop} shows that when $\alpha=0$, the behavior relatively follows the time-averaged case.
This is because the residual fluctuations are minimal for $\alpha = 0$.

\section{
Growth of Perturbations under the Kapitza-Type Modulations
}
\label{sec: particle production}

In the previous section, we discussed the spatially homogeneous components of the two axions.
In this section, we continue to treat the heavy axion $\phi$ as a spatially homogeneous classical background field,
and focus on the evolution of perturbations $\delta\chi$ of the light axion.

The perturbations can be quantized in Fourier space as
\begin{align}
    \delta\chi(t,\bm{x}) 
    = 
    \int \frac{\mathrm{d}^3k}{(2\pi)^{3/2}} 
    \left(
        \hat{a}_{\bm{k}}\chi_k(t) e^{-i\bm{k}\cdot \bm{x}}
        +
        \hat{a}^\dagger_{\bm{k}}\chi^\ast_k(t) e^{i\bm{k}\cdot \bm{x}}
    \right),
\end{align}
where $\hat{a}_{\bm{k}}$ and $\hat{a}^\dagger_{\bm{k}}$ are the annihilation and creation operators that satisfy the canonical commutation relations, and $\chi_k$ is the mode function.
To linear order,
the mode functions obey the equation of motion,
\begin{align}
    \ddot{\chi}_k+\left[k^2+R_m^2m_\phi^2\cos\left(N\frac{\phi}{f_\phi}+\frac{\bar{\chi}}{f_\chi}\right) \right]\chi_k
    =
    0,
    \label{eq: chik}
\end{align}
which describes a harmonic oscillator with a time-dependent angular frequency,
\begin{align}
    \omega_k^2(t)
    =
    k^2 + R_m^2 m_\phi^2 
    \cos\left(
        N\frac{\phi(t)}{f_\phi} + \frac{\bar{\chi}(t)}{f_\chi}
    \right),
    \label{eq: omega_k}
\end{align}
determined by the background fields $\phi(t)$ and $\bar{\chi}(t)$.
Due to the time dependence of $\omega_k^2$, $\chi_k$ can experience instabilities and grow exponentially as $\chi_k\propto\mathrm{exp}\left(\mu_k t /2 \right)$, where $\mu_k$ is the growth rate.
As the perturbations grow, they eventually backreact on the background fields, leading to the breakdown of the perturbative treatment. A larger growth rate makes this breakdown occur more rapidly.
In the following, we estimate the growth rate arising from parametric resonance and tachyonic effects, and discuss the condition under which the growth rate is suppressed. For definiteness, we assume the vacuum mode function 
as the initial condition for $\chi_k$,
\begin{equation}
\begin{aligned}
    \chi_k(0)
    &= \frac{1}{\sqrt{2 \omega_k(0)}}
    ,
    \\
    \dot{\chi}_k(0) &= - i \sqrt{\frac{\omega_k(0)}{2}},
\end{aligned}
\label{eq: chi_k initial}
\end{equation}
while our results are insensitive to this particular choice of initial condition.

\subsection{Parametric Resonance}

First, we focus on the parametric resonance of $\chi_k$ driven by the oscillations of the heavy axion $\phi(t)$.
In this case, the oscillations of $\phi$ make the effective frequency of $\chi_k$ time-dependent, leading to exponential growth of $\delta\chi_k$ through parametric resonance.
For simplicity, we consider the case where $J_0(A) < 0$ and $\bar{\chi}$ remains sufficiently close to the origin (see Fig.~\ref{fig: hilltop}).
In this case, the parametric resonance is most effective

If $\chi_k$ grows with a constant exponent, the growth rate is given by 
\begin{align}
    \mu_k
    =
    \frac{\log{|\chi_k\left(t_1\right)|^2}-\log{|\chi_k\left(t_2\right)|^2}}{t_1-t_2}.
    \label{eq: estimate growth rate}
\end{align}

The modulus of fluctuation $|\chi_k|$ oscillates and grows as $\bar{\chi}$ oscillates.
The growth rate is extracted from the change in the maximum amplitude of each oscillation.

\begin{figure}[t]
    \centering
    \includegraphics[width=0.7\linewidth]{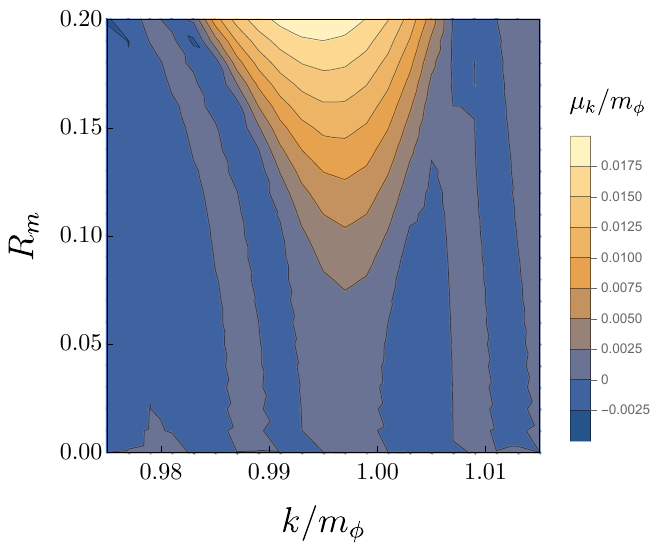}
    \caption{Numerical result for the growth rate with $\phi(0)=f_\phi$, $\bar{\chi}(0)\sim\pi f_\chi$, and $N=3$. An instability band is found around $k/m_\phi \simeq 1.00$.
    }
    \label{fig: narrow}
\end{figure}

Fig.~\ref{fig: narrow} shows the numerical results. We consider the case with $\bar{\chi}(0) \simeq \pi f_\chi$ and take $R_m \sim \mathcal{O}(0.1)$, which corresponds to a parameter choice near the boundary of the Kapitza regime.
In this marginal region, an instability band appears around $k/m_\phi \simeq 1.00$, and we numerically confirm that the growth rate $\mu_k$ scales as $R_m^2$.
For smaller $R_m$ within the Kapitza regime, however, the growth rate becomes too suppressed to be observed, and no instability band appears.

\subsection{Tachyonic instability}

Next, we consider the region where $k$ is much smaller than $m_\phi$. 
With the background solution of $\phi(t)$ given by \eqref{eq: phimotion}, $\chi_k$ oscillates with a time-dependent frequency 
\begin{align}
    \omega_k
    =
    \sqrt{
        k^2 + m_\chi^2 \cos\left(
            A \cos\left(m_\phi t+\alpha\right)
            +
            \frac{\bar{\chi}}{f_\chi}
        \right)
    }.
\end{align} 
For $k < m_\chi$, the angular frequency $\omega_k$ can be imaginary depending on the values of $\phi$ and $\bar{\chi}$, which indicates a tachyonic instability.
In particular, when $A$ is sufficiently large, $\omega_k$ oscillates between real and imaginary values following the oscillation of $\phi$.
Here, we assume 
$R_m \lesssim \mathcal{O}(0.01)$
so that the oscillation of $\bar{\chi}$ is much slower than that of $\phi$.
Then, we can approximately treat $\bar{\chi}$ as a constant and take the time average over the $\phi$ oscillation.
As a result, we obtain the equation of motion for $\chi_k$ as
\begin{align}
    \ddot{\chi}_k+\left[k^2+m_\chi^2J_0\left(A\right)\cos\left(\frac{\bar{\chi}}{f_\chi}\right)\right]\chi_k
    =
    0.
\end{align}
Thus, we roughly estimate the growth rate as
\begin{align}
    \mu_k
    \sim
    2\sqrt{-m_\phi^2R_m^2J_0(A)\cos \theta_\chi
    - k^2}.
    \label{eq: tachyon growth rate}
\end{align}
Since the growth rate depends linearly on $R_m$ for small $k$, its effect is more significant than parametric resonance for small $R_m$.

\begin{figure}[t]
    \centering
    \includegraphics[width=1\linewidth]{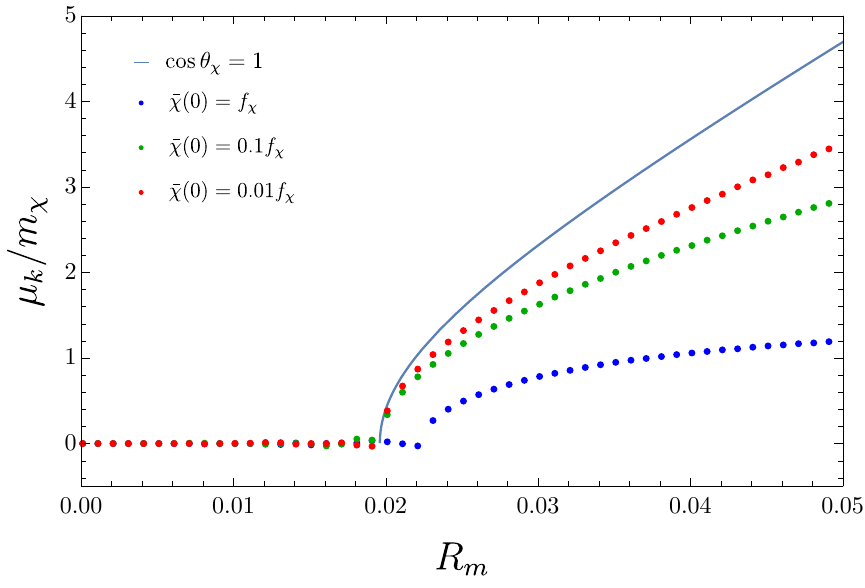}
    \caption{The numerical result of the growth rate with $A=3$ and $k=m_\chi$. 
    The colored points represent the results for different initial condition $\bar{\chi}(0)$.
    The solid line represents the analytical estimate Eq.~\eqref{eq: tachyon growth rate} for the case of $\cos \theta_\chi=1$, corresponding to the maximal growth.
   }
    \label{fig: growthrate}
\end{figure}

We show the growth rate for different $R_m$ and $\bar{\chi}(0)$ in Fig.~\ref{fig: growthrate}, where the growth rate is estimated in the same way as before, using Eq.~\eqref{eq: estimate growth rate}.
Here, we fix $k =  m_\chi$ and $A = 3$, corresponding to $J_0(A) \simeq -0.26$.
Then, we find nonzero growth only for $R_m > k/(\sqrt{|J_0(A)|} m_\phi) \simeq 0.02$.
Since $\bar{\chi}$ oscillates around $\pi f_\chi$ due to negative $J_0(A)$, a smaller $\bar{\chi}(0)$ leads to a larger amplitude of oscillations, which means that $\cos \theta_\chi$ takes positive values for a longer time.
As a result, the growth rate becomes larger for smaller $\bar{\chi}(0)$.
For comparison, we also show the analytic estimate~\eqref{eq: tachyon growth rate} with $\cos \theta_\chi = 1$ as the maximal growth by a solid line.
As expected, the actual growth rate is smaller than the maximal estimate.

\begin{figure}[t]
    \centering
    \includegraphics[width=0.7\linewidth]{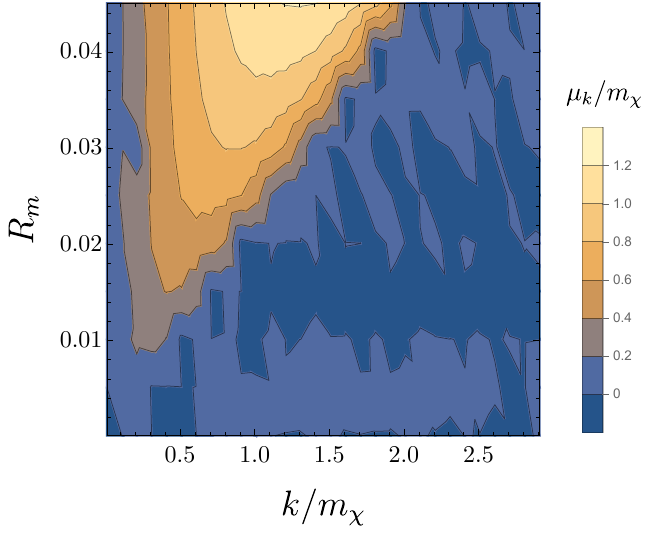}
     \caption{The growth rate of occupation number with $\Phi=f_\phi,N=3$ and $\bar{\chi}(0)=f_\chi$.
     }
    \label{fig: broad}
 \end{figure}

We show the numerical evaluation of the growth rate for different $k$ and $R_m$ in Fig.~\ref{fig: broad}.
Here, we fix $A = 3$ and $\bar{\chi}(0) = f_\chi$.
While Eq.~\eqref{eq: tachyon growth rate} suggests that the growth rate is maximized at $k=0$, numerical calculations show that it is actually suppressed for $k \sim 0$.
The actual evolution of $\chi_k$ is described by alternating phases of $\omega_k^2 > 0$ and $\omega_k^2 < 0$, which results in a complicated evolution of $\chi_k$.
Consequently, the actual growth rate in the vicinity of $k^2 \sim 0$ is not fully reproduced by Eq.~\eqref{eq: tachyon growth rate}, which was obtained from the time-averaged effective equation of motion, and $\chi_k$ does not efficiently grow for too small $k$.
On the other hand, the upper bound on $k$ of the instability band is well described by $k/m_\phi \lesssim \sqrt{|J_0(A)|} R_m$ as expected from Eq.~\eqref{eq: tachyon growth rate}.

Finally, we estimate the magnitude of the perturbations to check the validity of the perturbative approach. 
Now, we focus on small values of $R_m$,
where the time-averaging approximation is valid, and neglect the parametric resonance growth.
The growth of the perturbations is approximately given by
\begin{align}
    \langle(\delta\chi)^2\rangle
    =
    \int \mathrm{d}\ln k 
    \frac{k^3}{2 \pi^2}|\chi_k|^2
    =
    \int \mathrm{d} \ln k \, 
    k^3\frac{1}{4\pi^2\omega_k(0)}e^{ \mu_k t}.
\end{align} 
The perturbation grows when $\mu_k$ in Eq.~\eqref{eq: tachyon growth rate} takes real values. 
To obtain an order-of-magnitude estimate, let us evaluate the growth by setting $J_0(A)\cos\theta_\chi \sim -1$.
Then, the tachyonic instability typically occurs for $k \lesssim m_\phi R_m$, and the perturbation grows as

\begin{align}
    \langle\delta\chi^2\rangle
    \sim
    \int_0^{m_\chi} \mathrm{d} k \, k
    \exp\left(
         m_\chi t
    \right).
\end{align}
For reference, in the parameter choice adopted in Sec.~\ref{sec: app}, namely $m_\chi = 10^{-2}\,\mathrm{eV}$ and $f_\chi = 10^{11}\,\mathrm{GeV}$, the time required for the perturbation to grow to the same order as the background field is approximately
\begin{equation}
t \simeq {\cal O}(10^2)\, m_\chi^{-1}.
\end{equation}
Of course, this corresponds to the case where the tachyonic instability is maximized. This is only an order-of-magnitude estimate, and in practice the perturbative description is expected to remain valid over longer timescales. In particular, in an expanding universe the growth of perturbations is further suppressed because the oscillation amplitude decreases due to Hubble friction. When the oscillation amplitude becomes smaller than $\pi f_\chi/2$, the frequency $\omega_k$ no longer has an imaginary part, and the tachyonic instability does not occur.
Unless the system starts extremely close to the hilltop or under other special initial conditions, the backreaction can be safely neglected, for cosmological applications, for at least $\mathcal{O}(10^3)$ oscillation periods of $\chi$, or even longer.

\section{Cosmological Applications}
\label{sec: app}

So far, we have discussed the dynamics of axions under the Kapitza-type modulation in Minkowski spacetime. 
We now consider the expanding universe, focusing on the radiation-dominated epoch, and investigate the cosmological implications of the Kapitza-type modulation for axion dynamics.

When the time-averaging approximation is valid, the field $\bar{\chi}$ can be regarded as evolving in an effective potential (\ref{eq: effective potential}).
In the expanding universe, $A$ decreases as the amplitude of $\phi$ redshifts.
Consequently, the effective potential becomes time-dependent, with its stable points dynamically changing in time.
The effects of the Kapitza-type modulation are most significant while the heavy axion oscillates with a large amplitude, since the first zero of $J_0(A)$ occurs at $A \approx 2.4$.
For this modulation to affect the dynamics of the light axion, its mass should be larger than the Hubble parameter when $A$ is still large.
To realize such a situation while keeping a large mass hierarchy between the two axions, 
we consider the clockwork mechanism~\cite{Choi:2014rja,Choi:2015fiu,Kaplan:2015fuy,Higaki:2015jag,Higaki:2016yqk}. 
This yields $N \gg 1$ and therefore an initially large $A \gg 1$, 
so that $J_0(A)$ crosses zero many times and the light axion oscillates successively around different minima.\footnote{
Alternatively, we could adopt the delayed onset of the $\phi$ oscillations via the trapped misalignment mechanism~\cite{Higaki:2016yqk,Nakagawa:2020zjr,DiLuzio:2021gos,Jeong:2022kdr,DiLuzio:2024fyt}. In this case, the sign flip occurs only a few times for $N = {\cal O}(1)$.
}
Eventually, the light axion settles into oscillations around the true vacuum, $\bar{\chi} = 2 n \pi f_\chi$ for integer $n$, after $J_0(A)$ becomes positive for the last time.

\subsection{Analytical estimate of the axion abundance}

We now estimate the abundance of the light axion based on the effective potential (\ref{eq: effective potential}).
In the radiation-dominated era, the Hubble parameter and the scale factor are given by
\begin{align}
    H&=\frac{1}{2t},
    \quad 
    a(t)
    =
    \sqrt{\frac{t}{t_\mathrm{in}}},
\end{align}
where we normalize the scale factor so that $a(t_\mathrm{in}) = 1$, and the subscript `in' implies that the quantity is evaluated at the time when the Hubble parameter becomes comparable to the light axion mass, i.e., $3H(t_\mathrm{in}) = m_\chi$.

The oscillation of the heavy axion can be represented by Eq.~\eqref{eq: phimotion} with a time-dependent amplitude $\Phi(t)$, which decreases due to the cosmic expansion.
We define $\Phi_\mathrm{in} \equiv \Phi(t_\mathrm{in})$ and $A_\mathrm{in} \equiv A(t_\mathrm{in}) = N \Phi_\mathrm{in}/ f_\phi$.
The time evolution of $A$ is given by
\begin{align}
    A(t)
    =
    A_\mathrm{in} a(t)^{-\frac{3}{2}}
    =
    A_\mathrm{in} \left(\frac{t}{t_\mathrm{in}}\right)^{-\frac{3}{4}}.
    \label{eq: A}
\end{align}
for $t > t_\mathrm{in}$.
Here, we assumed that the evolution of $\Phi$ can be approximated by its dynamics in a quadratic potential.

As $A(t)$ decreases, the overall sign of the effective potential flips, and the light axion eventually starts to oscillate around the true vacuum at $\theta_\chi = 2n\pi$. 
We denote the onset time of this oscillation by $t_\mathrm{osc}$, which is determined by the condition 
$3H(t_\mathrm{osc}) = m_\mathrm{eff}(t_\mathrm{osc})$ and  
$ m_\mathrm{eff}(t) \equiv \sqrt{|J_0(A(t))|}\, m_\chi$,
subject to a condition that $A(t_\mathrm{osc})$ is smaller than the first zero of $J_0(A)$.

Now we focus on the dynamics in which the light axion, 
oscillating around a false vacuum due to the Kapitza effect, 
eventually starts oscillating around the true vacuum as the effective potential flips for the last time. 
Since $A$ decays on a Hubble timescale $H^{-1}$, the flip of the effective potential also occurs on this timescale. 
Although the Kapitza effect only requires the heavy axion to oscillate rapidly, here we consider the situation where the light axion also undergoes many oscillations within a Hubble time in the effective potential.

This requirement leads to the condition,
\begin{align}
    H_\mathrm{osc}
    &\ll 
    H_\mathrm{in},
\end{align}    
namely,
\begin{align}
    m_\mathrm{eff}(t_{\rm osc})
    &\ll
    m_\chi,
    \label{condition}
\end{align}
is around $A\simeq 2.4$, Eq.~\eqref{eq: A} gives 
\begin{align}
    t_\mathrm{osc}
    \simeq
    \left(\frac{A_\mathrm{in}}{2.4}\right)^\frac{4}{3}t_\mathrm{in}
    ,
\end{align}
and the effective mass is given by
\begin{align}
    m_\mathrm{eff}(t_\mathrm{osc})
    =
    3H(t_\mathrm{osc})
    \simeq
    \left(\frac{A_\mathrm{in}}{2.4}\right)^{-\frac{4}{3}}
    \frac{3}{2t_\mathrm{in}}
=
    \left(\frac{A_\mathrm{in}}{2.4}\right)^{-\frac{4}{3}} m_\chi.
\end{align}

Here we examine the motion of the light axion. The crucial point is the moment when the sign of $J_0(A)$ flips from negative to positive. Before this transition, the light axion is expected to oscillate around $(2n+1)\pi f_\chi$ ($n$ an integer). In particular, if $A_{\rm in} \gg 1$, the axion should have oscillated around that vacuum for at least a Hubble time, and its oscillation amplitude at the time of the sign flip can be expected to be of order ${\cal O}(0.1) f_\chi$. Once the sign becomes positive, the light axion starts to oscillate around one of the true vacua around $2n'\pi f_\chi$ ($n'$ an integer), but its initial amplitude at that stage is expected to be rather large. A rough estimate suggests $\theta_{\rm osc} \simeq \pi$, where $\theta_{\rm osc}$ is the oscillation amplitude at $t=t_{\rm osc}$. Of course, this depends on details such as the exact timing of the sign flip and the oscillation phase at that moment, and anharmonic effects may also become relevant, so this is only a crude estimate. Note that the axion mass increases after $t_{\rm osc}$, and the resulting abundance can be estimated in the same manner as for the QCD axion.

On the other hand, if $A_{\rm in} \simeq 1$, the value of $J_0(A_{\rm in})$ is already of order ${\cal O}(0.1)$ and not very small, so $t_{\rm osc}$ and $t_{\rm in}$ are close to each other. In this case the axion begins to oscillate around the true vacuum from the outset, and the initial  amplitude can be approximated as $\theta_{\rm osc} \sim 1$.

The axion number density $n_\chi$ at $t=t_{\rm osc}$ is given by 
\begin{align}
    n_\chi(t_\mathrm{osc})
    =\frac{1}{2}m_\mathrm{eff}(t_\mathrm{osc})f_\chi^2
\theta_\mathrm{osc}^2
\simeq
\frac{1}{2}\left(\frac{A_\mathrm{in}}{2.4}\right)^{-\frac{4}{3}} m_\chi f_\chi^2\theta_\mathrm{osc}^2.
\label{eq: number density}
\end{align}

The entropy density and the Hubble parameter are given in terms of temperature as follows,
\begin{gather}
    s=\frac{2\pi^2}{45}g_{\ast s}(T)T^3,
    \\
    H^2\simeq\frac{\pi^2}{90}g_\ast(T)\frac{T^4}{M_\mathrm{Pl}^2},
\end{gather}
where $M_\mathrm{Pl} \simeq 2.435 \times 10^{18}$\,GeV is the reduced Planck mass, and $g_\ast$ and $g_{\ast s}$ are the effective number of degrees of freedom for energy density and entropy density, respectively.
The ratio of the axion number density to the entropy density is a conserved quantity and given by
\begin{align}
    \frac{n_\chi}{s}\bigg|_0
    =
    \frac{n_\chi}{s}\bigg|_\mathrm{osc}
    \simeq\,
    0.94
    \times \frac{g_\ast\left(T_\mathrm{osc}\right)^{\frac{3}{4}}}{g_{\ast s}\left(T_\mathrm{osc}\right) }\left(\frac{A_\mathrm{in}}{2.4}\right)^{\frac{2}{3}}\theta_\mathrm{osc}^2\left(\frac{m_\chi}{10^{-2}\,\mathrm{eV}}\right)^{-\frac{1}{2}}\left(\frac{f_\chi}{10^{11}\,\mathrm{GeV}}\right)^{2},
    \label{eq: number entropy ratio}
\end{align}
where the subscript `$0$' denotes the present value,
and we expect $\theta_{\rm osc} \sim \pi$ when the condition \eqref{condition} is satisfied.
On the other hand, if $A_{\rm in} \lesssim 1$, we expect $\theta_{\rm osc} \sim 1$  and do not have the enhancement factor of $(A_\mathrm{in}/2.4)^{2/3}$.
Thus, the axion abundance is given by
\begin{align}
      \Omega_\chi h^2
    =&
    m_\chi \frac{n_\chi}{s}\bigg|_\mathrm{0}\frac{s_0}{\rho_\mathrm{crit}h^{-2}}\notag\\
    =&
    \begin{cases}
     \displaystyle{8.6\times10^{-4}\theta_\mathrm{osc}^2\left(\frac{g_\ast}{80}\right)^{-\frac{1}{4}}\left(\frac{m_\chi}{10^{-2}\mathrm{eV}}\right)^{\frac{1}{2}}\left(\frac{f_\chi}{10^{11}\mathrm{GeV}}\right)^{2}},
     &
     (A_\mathrm{in}<2.4)
     \\
    \displaystyle{8.6\times10^{-4}\theta_\mathrm{osc}^2 \left(\frac{A_\mathrm{in}}{2.4}\right)^\frac{2}{3} \left(\frac{g_\ast}{80}\right)^{-\frac{1}{4}}\left(\frac{m_\chi}{10^{-2}\mathrm{eV}}\right)^{\frac{1}{2}}\left(\frac{f_\chi}{10^{11}\mathrm{GeV}}\right)^{2}},
    &
    (A_\mathrm{in}>2.4)
    \end{cases}
    \label{analytical estimate}
\end{align}
where $\rho_\mathrm{crit}\simeq(0.003\,\mathrm{eV})^4h^2$ is the critical density, $s_0\simeq2.2\times10^{-38}\,\mathrm{GeV}^3$
and we assume $g_\ast(T_\mathrm{osc})\simeq g_{\ast s}(T_\mathrm{osc})$.

\begin{figure}[t]
    \includegraphics[width=0.85\linewidth]{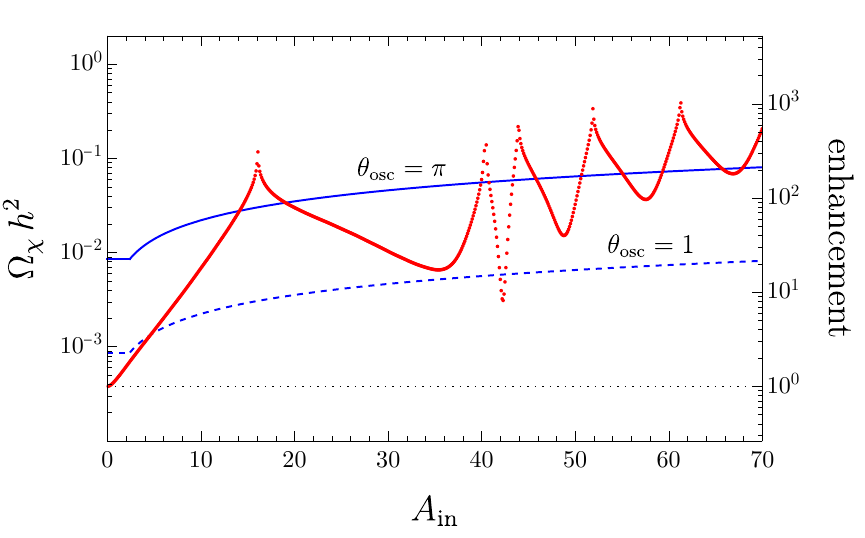}
    \caption{
    Axion abundance as a function of $A_{\rm in}$ for $m_\chi=10^{-2\,\mathrm{eV}}$, $f_\chi=10^{11\,\mathrm{GeV}}$, $R_m = 0.01$, $\alpha = 0$, and $g_\ast(T_\mathrm{osc})=80$ (see Eq.~\eqref{analytical estimate}). The initial conditions are set as $\bar\chi=f_\chi$ and $\dot{\bar{\chi}}=0$ at $t = 0.01m_\chi^{-1}$. The blue solid and dashed lines show the analytical estimates for $\theta_{\rm osc}=\pi$ and $\theta_{\rm osc}=1$, respectively, while the red dots represent the numerical results. The dotted horizontal line corresponds to the numerical result at $A_{\rm in}=0$ (without mixing), extended along the horizontal axis as a reference.}    
    \label{fig: enhance}
\end{figure}

We show the results of both numerical calculations and analytical estimate \eqref{analytical estimate} in Fig.~\ref{fig: enhance}. 
In our numerical calculation, the abundance is estimated at $t=500\,m_\chi^{-1}$, when the oscillation of $\phi$ has become sufficiently small and the effective mass has become 
approximately equal to $m_\chi$. The blue solid and dashed lines show the analytical estimates for $\theta_{\rm osc}=\pi$ and $\theta_{\rm osc}=1$, respectively. The red dots represent the numerical results. The dotted horizontal line corresponds to the numerical result at $A_{\rm in}=0$ (without mixing), extended along the horizontal axis, and the right-hand ticks show the enhancement of the abundance relative to this baseline. For large $A_{\rm in}$, the analytical estimate with $\theta_{\rm osc} = \pi$ and the numerical results are indeed in rough agreement, but the numerical results exhibit several characteristic spikes. These arise when, after the final sign flip, the axion happens to start oscillating near the hilltop, so that the anharmonic effect becomes significant. This anharmonic effect was not included in the analytical estimate. On the other hand, for small $A_{\rm in}$, the analytical result with $\theta_{\rm osc} = 1$ agrees well with the numerical estimate with $A_{\rm in} = 0$.
We can see that, compared to the dotted line,  the axion abundance can be enhanced by about two or three orders of magnitude due to the Kapitza effect.

So far we have neglected the backreaction of $\chi$ on the dynamics of $\phi$. For this to hold, the condition \eqref{eq: bacreaction} must be satisfied until the onset time of the last oscillation $t_\mathrm{osc}$:
\begin{align}
m_\phi^2 f_\phi \Phi_{\rm osc} \gg N m_\chi^2 f_\chi^2,
\label{eq: cond}
\end{align}
where $\Phi_{\rm osc}$ denotes the oscillation amplitude of $\phi$ at $t = t_{\rm osc}$.
The ratio of the energy densities of the two axions at the present time is estimated as
\begin{align}
\left. \frac{\rho_\phi}{\rho_\chi} \right|_0
\sim
\frac{m_\phi^2 \Phi_{\rm osc}^2}{\left(\frac{A_\mathrm{in}}{2.4}\right)^{-\frac{4}{3}} m_\chi^2 f_\chi^2 \theta_\mathrm{osc}^2}
\gg
\frac{N \Phi_{\rm osc}}{\left(\frac{A_\mathrm{in}}{2.4}\right)^{-\frac{4}{3}} f_\phi \theta_\mathrm{osc}^2}
\gtrsim
\left(\frac{A_\mathrm{in}}{2.4}\right)^{\frac{4}{3}} \frac{1}{\pi^2},
\end{align}
where the first inequality follows from the condition \eqref{eq: cond}, and the second inequality uses $A_{\rm osc} = N \Phi_{\rm osc}/f_\phi\gtrsim 1$ (for the Kapitza effect to be relevant) and $\theta_{\rm osc}\sim \pi$.
Thus, the abundance of the heavy axion is generically larger than or comparable to that of the light axion whenever the Kapitza effect plays an important role. In particular, if $A_{\rm in}\gg 1$, the heavy axion abundance dominates. On the other hand, if the mass of $\phi$ is sufficiently large, it could be unstable and decay into massless or light degrees of freedom. In that case, the light axion $\chi$ can be the dominant component of dark matter. If $\phi$ is stable on cosmological time scales, both axions contribute to dark matter. We note that, if one is interested in scenarios where the two axions have comparable abundances or $\chi$ dominates dark matter, the backreaction on $\phi$ must be taken into account. We leave this case for future work.

Finally, we comment on possible impacts of the Kapitza effect on isocurvature perturbations of the light axion. 
Even when the growth of perturbations discussed in Sec.~\ref{sec: particle production} is negligible, superhorizon fluctuations at $t \sim t_\mathrm{osc}$ effectively shift the initial value of $\bar{\chi}$ in each Hubble patch, modifying the subsequent dynamics of the light axion.
In particular, this effect is significant for $A_\mathrm{in}$ near the spikes in Fig.~\ref{fig: enhance}.
For such $A_\mathrm{in}$, the light axion approaches the hilltop after the final sign flip, and thus small shift of the initial condition results in a large isocurvature fluctuations of the light axion abundance. 
In an extreme case it can even  change the true vacuum around which the light axion finally oscillates, leading to the formation of domain walls.

\subsection{Applications to the QCD axion}

In the conventional trapped misalignment mechanism, the axion is trapped in a false vacuum by introducing an explicit PQ symmetry-breaking potential, which delays the onset of its oscillations.
However, due to the strong CP problem, the axion field must settle near zero. As a result, the CP-violating term requires a tuning at the level of $10^{-3}$~\cite{Jeong_2022}.
On the other hand, instead of introducing an explicit PQ symmetry-breaking potential, we assume a mixing with a heavier axion. 
In this case, due to the Kapitza effect, the effective potential for the QCD axion can be flipped. 
As the QCD potential is generated, the axion field is directed toward a false vacuum that does not coincide with the minimum of the original potential. 
Later, as the amplitude of the heavy axion becomes suppressed, the QCD potential returns to its original form, and the axion begins to oscillate around the true vacuum.
This mechanism effectively delays the onset of oscillations around the true vacuum, similar to the conventional trapped misalignment scenario.
The key point of this scenario is that the delay in the onset of QCD axion oscillation is caused by the flipping of the effective potential.
As a result, the vacuum is always located at the CP-conserving point (i.e., zero), and the strong CP problem remains solved. 
In other words, the Kapitza effect allows us to significantly enhance the axion abundance while avoiding the fine-tuning associated with the strong CP problem.

\section{Discussion and Conclusions}
\label{sec: summary}

So far, we have seen that the effective potential for the light axion, obtained by time-averaging the oscillations of the heavy axion, depends on the amplitude of the latter. In a cosmological setting,  the amplitude of the heavy axion decreases as $a^{-3/2}$ under the harmonic approximation. Therefore, a large initial value of $A$ is required for the effective potential to have a significant impact on the dynamics of the light axion. 
One possibility is to have a large mixing parameter $N \gg 1$,
which can naturally arise in the clockwork mechanism~\cite{Choi:2014rja,Choi:2015fiu,Kaplan:2015fuy,Higaki:2015jag,Higaki:2016yqk,Farina:2016tgd,Long:2018nsl}.
Alternatively, we could delay the onset of oscillations of the heavy axion by the trapped misalignment~\cite{Higaki:2016yqk,Nakagawa:2020zjr,DiLuzio:2021gos,Jeong:2022kdr,DiLuzio:2024fyt} or bubble misalignment~\cite{Nakagawa:2022wwm,Lee:2024oaz} mechanisms.

If the initial value of $A$ at the onset of heavy axion oscillations is much greater than unity, the effective potential for the light axion will flip its sign multiple times during the evolution of the universe. In particular, when the potential temporarily vanishes, the light axion is expected to move with an approximately constant velocity, passing over many potential maxima and minima until the potential grows again. This behavior is likely analogous to the axion roulette phenomenon observed in level-crossing scenarios~\cite{Daido:2015bva,Daido:2015cba}.

It is also conceivable that the dynamics of the heavy axion itself is governed by an effective potential generated by time-averaging the oscillations of an even heavier axion. In such a case, the amplitude of the heavy axion may remain large until relatively late times. A detailed study of the dynamics of such systems with the clockwork mechanism and multiple heavy axions having hierarchical masses and mixings is an interesting direction for future work.

In the previous section, we have focused on the enhancement of axion dark matter via this Kapitza-like effect. Another interesting possibility arises when the light axion starts from a hilltop initial condition and rolls in a single direction, as shown in Fig.~\ref{fig: hilltop}. This behavior can also occur when the effective potential nearly vanishes, as discussed above. Such dynamics could have applications to spontaneous baryogenesis~\cite{Cohen:1987vi,Cohen:1988kt,Dine:1990fj,Cohen:1991iu,Dolgov:1994zq,Dolgov:1996qq,Chiba:2003vp,Takahashi:2003db} or the formation of domain walls. A more detailed investigation of these possibilities is also left for future work.

In this paper, we have studied a two-axion system with hierarchical masses and explored how the rapid oscillations of the heavy axion dynamically change the effective potential of the light axion via a Kapitza-type mechanism. In particular, we demonstrated that the effective potential for the light axion can flip its sign or be nearly canceled, depending on the amplitude of the heavy axion oscillations.

We have numerically verified both the validity and the limitations of the time-averaged effective potential. As a concrete application, we considered a scenario in which the heavy axion modulates the potential of the light axion. To achieve a sufficiently large oscillation amplitude of the heavy axion at late times, we assumed a trapped or bubble misalignment mechanism. Under these conditions, we estimated the resulting light axion abundance and found that it is significantly enhanced compared to that from the standard misalignment mechanism. 

In the Kapitza-like dynamics, the light axion can undergo a large field excursion as seen in Fig.~\ref{fig: discrepancy}.
If it couples to photons, such a dynamics can induce isotropic cosmic birefringence, which is hinted by recent analyses of the cosmic microwave background (CMB) polarization~\cite{Minami:2020odp,Diego-Palazuelos:2022dsq,Eskilt:2022wav,Eskilt:2022cff,Cosmoglobe:2023pgf}.
To induce isotropic birefringence in the CMB polarization, the light axion must evolve after the recombination epoch.
In this case, the light axion must be a negligible or at least subdominant component of dark matter, while the heavy axion that induces the Kapitza effect can contribute more significantly to the dark matter abundance.

We emphasize that this work presents the first explicit application of Kapitza-type modulation in axion dynamics. The key ingredients for this phenomenon, namely axion mixing and a hierarchical mass spectrum, naturally arise in the axiverse. Therefore, the sign flip of the axion potential induced by such dynamical modulation may be a generic feature in multi-axion systems, with potentially significant implications for axion cosmology and dark matter physics.

\section*{Acknowledgments}
This work is supported by JSPS Core-to-Core Program (grant number: JPJSCCA20200002) (F.T.), JSPS KAKENHI Grant Numbers 20H01894 (F.T.), 20H05851 (F.T.), 23KJ0088 (K.M.), and 24K17039 (K.M.).
This article is based upon work from COST Action COSMIC WISPers CA21106, supported by COST (European Cooperation in Science and Technology).

\bibliographystyle{apsrev4-1}
\bibliography{ref}

\end{document}